\documentclass[aps,twocolumn,nofootinbib,longbibliography,showpacs]{revtex4-1}
\usepackage[babel]{csquotes}
\usepackage{graphicx}
\usepackage{amsmath,amssymb}
\usepackage[colorlinks,citecolor=blue,linkcolor=blue,urlcolor=blue]{hyperref}
\usepackage{mathrsfs}
\usepackage{enumerate}
\usepackage[toc,page]{appendix}
\def\be{\begin{equation}}
\def\ee{\end{equation}}

\newcommand{\dd}{ \mathrm{d}}

\newcommand{\beq}{\begin{equation}}
\newcommand{\eeq}{\end{equation}}

\newcommand{\Sch}{Schwarzschild }

\begin{document}

\title{How big is a black hole?}

\author{Marios Christodoulou and Carlo Rovelli}

\affiliation{\vspace{1mm}\mbox{CPT, Aix-Marseille Universit\'e, Universit\'e de Toulon, CNRS,}\\ 
\mbox{Samy Maroun Center for Time, Space and the Quantum.}\\ 
Case 907, F-13288 Marseille, France.}

\date{\small\today}

\begin{abstract}
\noindent 
The 3d volume inside a spherical black hole can be defined by extending an intrinsic flat-spacetime characterization of the volume inside a 2-sphere. For a collapsed object, the volume grows with time since the collapse, reaching a simple asymptotic form, which has a compelling geometrical interpretation. Perhaps surprising, it is large.  The result may have relevance for the discussion on the information paradox.
\end{abstract}


\maketitle
    
\section{Introduction}

How much space is there inside the black hole formed by a collapsed star of mass $m$?  Not too much, one might think: in flat space, the volume inside a sphere with radius $r\!=\!2m$ (in $G\!=\!c\!=\!1$ units) is $\frac43 \pi (2m)^3$: a few $km^3$ for a stellar black hole. But flat space intuition does not apply to the curved geometry inside the hole: inside an eternal hole described by Kruskal geometry, there is, in a  sense, an entire second asymptotic region. 

In fact, the question is not well posed: what do we mean by ``the" volume inside the horizon? Which 3d spacelike surface are we considering? The volume of the $t=const.$ surfaces, where $t$ is a time coordinate, depends on the arbitrary choice of coordinates.  The issue has been discussed by various authors \cite{DiNunno2009, Finch2012, Cvetic, Gibbons2012, Parikh2006, Grumiller, Ballik2010, Ballik2013}. 

\begin{figure}[b]
\vspace{-2mm}
\centerline{\includegraphics[height=5cm]{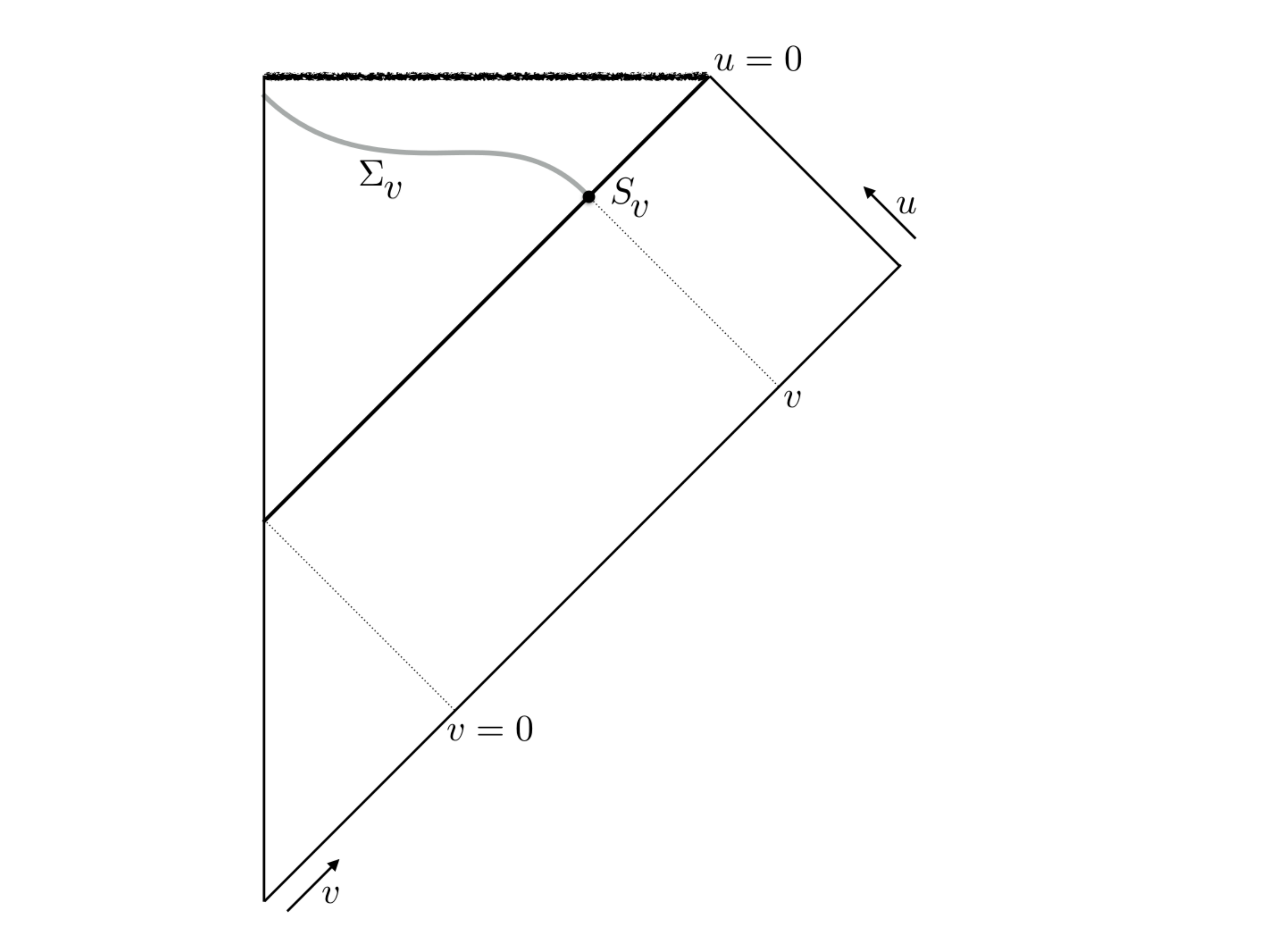}}
\caption{Conformal diagram of a collapsing object spacetime. The sphere $S_v$ is on the horizon, at time $v$. The spacelike surface $\Sigma_v$ whose volume we are computing is the one of maximum volume among those bounded by $S_v$.}
\label{fig:Figura1}
\end{figure}

Here we suggest a different way of thinking about the volume of the space inside a black hole. Our starting point is the simple observation that the {exterior} of the Schwarzschild metric is static, but the {interior} is not. The interior keeps changing. Therefore a good notion of (interior) volume can be time-dependent. The horizon is naturally foliated by two-spheres, and we can ask if there is a natural definition of ``interior volume" associated to a single two-sphere. 

In Minkowski spacetime there is a simple characterisation of the volume inside a two-sphere that remains meaningful in a spherically symmetric curved geometry: the volume inside a two-sphere $S$ is the volume of the \emph{largest} spacelike spherically-symmetric surface $\Sigma$ bounded by $S$.  This is what we mean by ``volume inside a sphere" in flat spacetime; $\Sigma$, indeed, lies on the simultaneity surface determined by $S$.  This characterisation provides a coordinate independent definition to the notion of ``volume inside a sphere" which remains valid in the case of spherical black holes, and captures the idea of ``how much space is inside". 

The horizon of a spherically symmetric hole is foliated by (spacelike) spheres $S_v$.  A convenient labelling of the spheres is asymptotic time, namely the null coordinate $v$ which at past infinity is related to Minkowski polar coordinates by $v=r+t$.  See Fig.~\ref{fig:Figura1}.  We find that the volume $V(v)$ inside the sphere $S_v$ grows with $v$. This makes sense: even if its surface-area remains constant, the horizon is still an outgoing null surface and the interior volume keeps growing with time. Matter, so to say, has newer and newer space where to fall into. 

In this paper we compute $V(v)$. The calculation demands solving the differential equation that determines the maximal-volume surface $\Sigma$.  We find that, setting $v\!=\!0$ at collapse time (see Fig.~\ref{fig:Figura1}), the volume takes the simple expression 
\be
V(v)  \underset{v\to\infty}{=} 3\sqrt{3}\,\pi\, m^2\, v 
\label{risultato}
\ee  
when $v$ is large with respect to $m$. The bulk of the volume turns out to be due to a region in the vicinity of a constant value of the radial coordinate. That is: inside the hole there is a long spacelike 3d cylinder with slowly varying radius, which grows longer with time.  

This is a surprising result, because the volume is \emph{large}.  For instance, the black hole Sagitarius A${^*}$ has radius $\sim10^6 km$ and age  $\sim 10^9 years$. Inside it, there is space for $\sim 10^{34} km^3$, enough to fit a million Solar Systems!

If a black hole of initial mass $m$ has a lifetime $\sim\!m^3$ in Planck units as predicted by Hawking radiation theory, there might be room inside it for a spacelike surface with volume $V\!\sim m^5$. For a stellar black hole, this is larger than our universe.  

There is a lot of available real estate inside a black hole, according to classical general relativity!

The result  can be extended to other spherically symmetric geometries, like the non-singular black hole metric considered in \cite{Hayward2005} and \cite{Rovelli2014}; in Appendix  \ref{app:RN} we treat the Reissner-Nordstr{\"o}m case. It can also be extended to the case of a Kruskal black hole. We do so in Appendix  \ref{app:KR}. The volume turns out to be infinite (as expected) but the difference $V(\Delta v)=V(v_2)-V(v_1)$ can be appropriately defined and is finite: it grows linearly in $\Delta v=v_2-v_1$ when $\Delta v$ is large.  

In the last section, we present some considerations on the relevance of these results for the ``information paradox" discussion. 

\section{Volume inside a sphere} \label{sec:sphereVolume}

Consider a (metric) 2d sphere $S$ immersed in flat Minkoswki spacetime. Let $R$ be its radius and $A=\pi R^2$ its area. We say that it encloses the volume $V=\frac43\pi R^3$. What does this mean?  It means that there is a 3d spacelike surface $\Sigma$ bounded by $S$ which has volume $V$. But there are a lot of spacelike surfaces bounded by $S$ in spacetime: which one do we mean, to define the interior volume? The answer can be given in two equivalent manners:
\begin{enumerate}
\item[(i)] $\Sigma$ lies on the same simultaneity surface as $S$.
\item[(ii)] $\Sigma$ is the largest spherically symmetric surface bounded by $S$.
\end{enumerate}
These two characterizations of $\Sigma$ are equivalent. To see this, we can chose (without loss of generality) coordinates $(x,y,z,t)$ where $S$ is given by $t=0, r^2\equiv x^2+y^2+z^2=R^2$. A spherically symmetric surface $\Sigma$ bounded by $S$ is defined by the function $t=t(r), r\in[0,R]$, with $t(R)=0$. Its volume is 
\be
V=\int_0^R dr\ 4\pi r^2\  \sqrt{1-\left(\frac{dt(r)}{dr}\right)^2}
\ee
which is maximised by $t(r)=0$ (because any variation adds a contribution in the timelike directions and reduces the volume), namely by the $\Sigma$ on the simultaneity surface.   The ``space inside $S$" is therefore the largest spherically symmetric space bounded by $S$. 

Let us now move to a curved spacetime. Given a sphere $S$ in a spherically symmetric geometry, what is the volume inside it?  Lacking flatness, simultaneity surfaces have no special significance, in general. But the second definition of the space inside $S$ extends immediately, and we adopt it here from now.  This allows us to define the volume $V$ enclosed in any 2d sphere $S$ in a spherically symmetric spacetime.

\section{Formulation of the problem}
Consider  the geometry of a collapsed object.  We work in ingoing Eddington-Finkelstein coordinates $(v,r,\theta,\phi)$. 
For simplicity, we take a null spherical shell of energy $m$ collapsing along the $v=0$ surface. Before this surface, spacetime is flat. After this surface, the geometry is a  \Sch black hole and the line element is the standard  \Sch geometry in  Eddington-Finkelstein coordinates 
\begin{equation}\label{eq:iEFlinel}
\dd s^2= -f(r)\dd v^2+  2 \dd v \dd r +r^2 \dd \Omega ^2,
\end{equation}
where $f(r)=1-2m/r$ and $\dd \Omega ^2 =  \sin ^2 \! \theta \: \dd \phi ^2 + \dd \theta^2$. The relation with the  \Sch  coordinates $t,r,\theta,\phi$, is given by  $v = t+\int\frac{dr}{f(r)}= t+r+2m\ln{\left|r-2m\right|}$.

The horizon is at $r=2m$. It is foliated by spheres $S_v$ defined by $r=2m$ and constant $v$.  The sphere $S_v$ is defined physically as the one crossed by a light signal sent by a stationary observer at large (with respect to $m$) distance $r$ from the hole, at proper time $t=v-r$. 

For each $S_v$ we are interested in the spherically symmetric 3d surface $\Sigma_v$ which is bounded by $S_v$ and has maximal 3d volume. The volume of this surface is called $V(v)$. Our objective is now to compute $V(v)$. This a well defined problem. 

\section{The maximisation problem} \label{sec:Volume}

A 3d spherically symmetric surface $\Sigma$, can be thought of as the direct product of a 2-sphere and a curve $\gamma$ in the $v$-$r$ plane.
\begin{eqnarray}\label{eq:sigmalambda}
\Sigma &\equiv & \gamma \times S^2\\
\gamma &\mapsto & (v(\lambda),r(\lambda))   .
\end{eqnarray}
The curve $\gamma$ is given here in parametric form, with an arbitrary parameter $\lambda$. We choose $\lambda=0$ on the horizon ($r=2m$) and call $\lambda_f$  ($f$ for `final') the value of $\lambda$ at $r=0$. Thus, the initial and final endpoints of $\gamma$ are given by
\begin{eqnarray}\label{eq:boundaryGamma}
&\ \   r(0)=2m, \hspace{2em}&  r(\lambda_f)=0,\\ 
& v (0) = v,  \hspace{2em}&  v(\lambda_f) = v_f .
\end{eqnarray} 

The surface $\Sigma$ is coordinatized by $\lambda,\theta,\phi$. The line element of the induced metric on $\Sigma$ is 
\begin{equation}\label{eq:sigmalinel}
\dd s^2_\Sigma= \left(-f(r)\dot{v}^2+  2 \dot{v} \dot{r} \right)\dd \lambda^2+r^2 \dd \Omega ^2
\end{equation}
where the dot indicates differentiation by $\lambda$. We hereafter do not consider the flat part inside the horizon (under the null shell) which would have contributed to the volume by about $4 \pi (2m)^3/3$. We are interested in the asymptotic behavior of the volume, when $v>>m$ and it will be seen that this contibution becomes negligible. The condition that $\Sigma$ is spacelike reads 
\beq \label{eq:spacelikeSigma}
-f(r)\dot{v}^2+  2 \dot{v} \dot{r}>0
\eeq
from the requirement that $\dd s^2|_\Sigma>0$ for all coordinate values. The proper volume of $\Sigma$ is given by
\begin{eqnarray}  
V_\Sigma [\gamma] &=& \int_0^{\lambda_{f}} \dd \lambda \,  \int_{S^2} \dd \Omega \,  \sqrt{r^4(-f(r)\dot{v}^2 + 2\dot{v}\dot{r})\sin^2{\theta}} \nonumber \\
&=& 4\pi \int _0 ^{\lambda_{f}} \dd \lambda \, \sqrt{r^4(-f(r)\dot{v}^2 + 2\dot{v}\dot{r})} \label{eq:volSigma}.
\end{eqnarray}

The surface $\Sigma_v$ that extremizes the volume is  determined by the curve $\gamma_v$ that extremizes this integral and by the specification of $v$ and $v_f$.   

\section{Geodesics in the auxiliary manifold}		
The last equation shows that finding $\Sigma_v$ is the same as solving for the equations of motion with langrangian 
\beq \label{eq:langrangian}
L(r,v,\dot{r},\dot{v})=\sqrt{r^4(-f(r)\dot{v}^2 + 2\dot{v}\dot{r})} . 
\eeq  
It is useful to note that this can be rewritten as 
\beq \label{eq:langrangianMaux}
L(r,v,\dot{r},\dot{v})=\sqrt{\tilde{g}_{\alpha \beta}\dd x^\alpha \dd x^\beta}
\eeq 
and can be thought of as the line element of an (auxiliary) 2d curved spacetime. That is, the metric $\tilde{g}_{\alpha \beta}$ is given by
\beq \label{eq:Mauxlinel}
ds^2_{M_{aux}} = r^4(-f(r)\dd v^2+  2 \dd v \dd r)
\eeq
where $\alpha,\beta,...$ can take the two values $v$ and $r$.  Finding $\Sigma_v$ is equivalent to finding the geodesics of this auxiliary  metric. 

Furthermore, equation \eqref{eq:volSigma} shows that the proper length of the geodesic in the auxiliary metric (times $4\pi$) is precisely  the volume of $\Sigma$. 

The condition \eqref{eq:spacelikeSigma} that $\Sigma$ be spacelike, suggests that $L > 0$, since $r$ is positive. The Langrangian appears to vanish at $r=0$, which is the final point for the geodesic but, as can be seen from \eqref{eq:rdoteq}, $\dot{r}$ becomes infinite.  Thus, $\gamma$ is a spacelike geodesic in $M_{aux}$. We now recognize that a well suited parametrization is to take $\lambda$ as the proper length in $M_{aux}$. After the extremization, we set 
\begin{eqnarray} \label{eq:lambdaPropLen}
L(r,v,\dot{r},\dot{v})&=&1 \nonumber \\ \nonumber \\
\Rightarrow  r^4(-f(r)\dot{v}^2 + 2\dot{v}\dot{r})&=&1
\end{eqnarray}
and from \eqref{eq:volSigma} we have immediately that  
\be \label{eq:volume}
V=4\pi \lambda_f.
\ee

The metric $\tilde{g}_{\alpha \beta}$ has a Killing vector, $\xi^\mu=(\partial_v)^\mu \propto (1,0)$. Since $\gamma$ is an affinely parametrized geodesic in $M_{aux}$, the inner product of $\xi$ with its tangent $\dot{x}^\alpha=(\dot{v},\dot{r})$, is conserved
\beq \label{eq:killingConstant}
r^4(-f(r)\dot{v} + \dot{r})=A
\eeq   

Equations \eqref{eq:lambdaPropLen} and \eqref{eq:killingConstant} are all we need to analyse the geodesics. They can be  recast in the form\footnote{The plus sign choice in \eqref{eq:rdoteq} would correspond to geodesics outside the horizon}
\begin{eqnarray}
\dot{r}&=& - \, r^{-4}\sqrt{A^2+r^4 f(r)} \label{eq:rdoteq}\\
\dot{v}&=&\frac{1}{A+r^4 \dot{r}} \label{eq:vdoteq}
\end{eqnarray}
It can be easily seen that $A$ has to be negative for the geodesic to be spacelike. Then, $\dot{v}$ and $\dot{r}$ are both negative and there are only positive terms in \eqref{eq:langrangian}. Integrating \eqref{eq:rdoteq} we get
\beq
\label{eq:rdotIntegral}
\frac{V_\Sigma}{4\pi} =\lambda_{f}= \int _{0} ^{2M} \dd r \; \frac{r^{4}}{\sqrt{A^2+r^4 f(r)}}. 
\eeq

Equation \eqref{eq:rdotIntegral} shows that there is a restriction imposed on $A$
\begin{eqnarray} \label{eq:Acritical}
A^2&+&r^4f(r)>0 \nonumber \\
\Rightarrow A^2&>-&r_V^4f(r_V)=\frac{27}{16} m^4\equiv A_c^2
\end{eqnarray}
The last condition comes from inspecting the polynomial $-r^4f(r)$. It has roots at $r=0$ and $r=2M$, is otherwise positive in that range and reaches a maximum at $r_V=\frac{3}{2}m$.  

We integrate numerically equations \eqref{eq:rdoteq} and \eqref{eq:vdoteq} below in Section \ref{sec:NumAnalysis}; but we can already directly derive the essential lesson by noticing the following: inserting a constant radial value for $r$, \eqref{eq:rdoteq} and \eqref{eq:vdoteq} become
\begin{eqnarray} \label{eq:rconstgeod}
A^2&=&-r^4 f(r) \\
\dot{v} \ &=&\frac{1}{A}  
\end{eqnarray}

Since $-r^4 f(r)>0$ in the range $0<r<2M$,  every constant $r$ provides a solution. In other words, the $r=const.$ surfaces are spacelike geodesics of the auxilliary manifold or equivalently stationary (maximal) points of the volume functional \eqref{eq:volSigma}. Integrating the $\dot{v}$ equation we get 
\beq \label{eq:lambdarconst}
 \lambda_f= A (v_f -v). 
\eeq
Thus, the $r=const.$ surface with the largest volume betwen two given $v$ is when $A$ is largest. That is, for $r=r_V$ which gives $A=A_c$. These considerations provide the basis for the  derivation of the asymptotic volume. This is done in the next section.

\section{Asymptotic expression for the volume}		
We are now interested in the volume for large $v$. The proper length of the geodesic should increase monotonically from $S$ which is situated at the endpoint $(v,2m)$ up to $(v_f,0)$. The $v$ coordinate of the point where $\Sigma_v$ reaches $r=0$ can be easily estimated: it must be before the formation of the singularity, because this increases the available volume, and in the large $v$ limit we can take $v_f=0$ without significative error. Let us therefore take the end of $\gamma$ at the coordinates $(0,0)$. 

Can we guess which path maximizes the volume? The crucial observation is that in order to maximize $\lambda_f$ when $v$ is very large, the geodesic must spent the maximum possible time at the radius $r$ where the line element is longer and the line element happens to have a maximum\footnote{Recall that the $\dot{r}=0$ surfaces extremize the volume, see end of previous section}. Therefore we may approximate the geodesic with an initial and a final transients and an intermediate long steady phase where $\dot r\sim 0$. Then the auxilliary line element \eqref{eq:Mauxlinel} becomes
\be \label{eq:rconstantLinel}
ds_{M_{aux}}\sim  -\sqrt{-r^4 f(r)} dv, 
\ee
with the approximation improving as $v$ incrases. The choice of the minus sign is needed because $\dd v<0$. To maximise the length, the steady phase of the geodesic must run at the value of $r$ that maximises $\dd s/\dd v$, which is given by 
\be \label{eq:rVcondition}
\frac{d}{dr} \sqrt{-r^4 f(r)}=0.
\ee
This value of $r$ is the one that maximizes the polynomial $-r^4 f(r)$, which we already called $r_V$
\be \label{eq:volumeRadius}
r_V=\frac32m 
\ee
Therefore, for large $v$ the largest spherically symmetric spacelike surface is formed by a long stretch at nearly constant radius $r_V=\frac32m$, joined to the $r=2M$ horizon on one side and to $r=0$ to the opposite extreme by  transients\footnote{We thank an anonymous referee for pointing out that the existence of the maximal slide at $r=\frac32m$, which is a key point of this paper, was already noticed numerically in \cite{Estabrook1973}, starting from a somewhat different approach.}. The infinitesimal proper length in the auxiliary metric is given by \eqref{eq:rconstantLinel}. Thus,
\begin{eqnarray} \label{eq:volEst}
V&\approx& -4\pi \, r_V^4 f(r_V)\, v \\ \nonumber
&=& - 4\pi \,A_c\, v\\
&=& 3 \sqrt{3}\,\pi\, m^2\, v \nonumber
\end{eqnarray}
which is the result anticipated in the introduction. It  is simply the combination of equations \eqref{eq:volume} and  \eqref{eq:lambdarconst} for $v_f=0$.

The result extends immediately to other spherically symmetric spacetimes defined by the metric  \eqref{eq:iEFlinel}, with a different function $f(r)$.  It is sufficient to find the maximum of  $-r^4 f(r)$ and the asymptotic expression for the volume is given by \eqref{eq:volEst} with $A_c =-\sqrt{-r_V^4 f(r_V)}$. 

\section{Numerical analysis} \label{sec:NumAnalysis}

\begin{figure}[t]
\vspace{.5cm}
 \centering
 \includegraphics[scale=0.4]{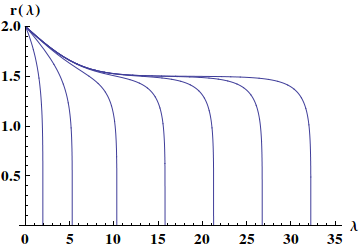}
 \caption{\label{fig:roflambda} The area coordinate $r$, in $m=1$ units, as a function of the volume parameter $\lambda$, obtained integrating the equation \eqref{eq:rdoteq}. As $A \rightarrow A_c$ , $\lambda_f=V / 4\pi \rightarrow \infty $.}
\end{figure}

\begin{figure}[t] 
 \centering
 \includegraphics[scale=0.30]{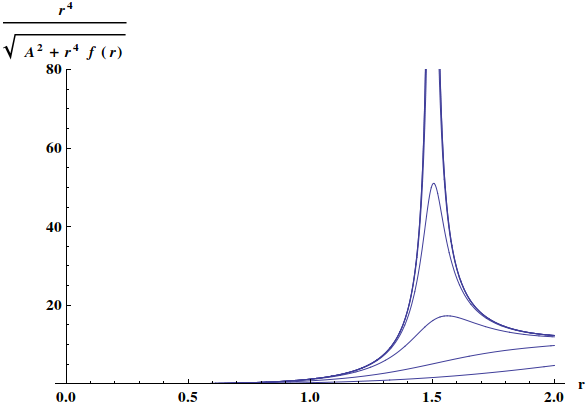}
 \caption{ \label{fig:integrand} The integrand in \eqref{eq:rdotIntegral} for different values of $A$ gradually approaching $A_c$. As $A\rightarrow A_c$ the volume contribution comes increasingly from  $r_V=3m/2$.}
\end{figure}

\begin{figure}[b]
 \centering
 \includegraphics[scale=0.4]{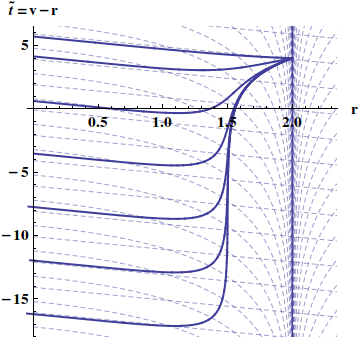}
 \caption{ \label{fig:maxGeod} Black hole spacetime in Eddington-Finkelstein coordinates. The horizon is the vertical line $r=2m$. 
 Dashed lines are the null geodesics.  Maximum volume surfaces for different values of $A$, starting from the same sphere on the horizon are depicted. }
\end{figure}

In order to verify the legitimacy of the approximations taken in the previous section, and to study the volume for finite times, we solve 
the equations defining $\Sigma_v$ numerically.  More precisely, we solve \eqref{eq:rdoteq} numerically and plot $r(\lambda)$ against $\lambda$. The volume is given by ($4\pi$ times) $\lambda_f$ which is such that $r(\lambda_f)=0$. The result of the numerical integration is given in Figure \ref{fig:roflambda} for a range of values of the integration constant $A$, which in turn determines $v$. In all Figures, we have plotted the values for $A^2=A_c^2+\{10,1,10^{-1},\ldots,10^{-5}\}$. 

The plot shows that the total volume increases as $A$ approaches $A_c$. Values of $A$ close to $A_c$ correspond to larger $v$. Figure  \ref{fig:roflambda} confirms the analysis of the previous section: for large volumes, the surface $\Sigma$ has two transient regions at the beginning and at the end, and a long steady region, where most of the volume builds up precisely at the value 1.5 (that is $\frac32 m$ in $m=1$ units) of the radius.  

To emphasise this point, we have plotted the integrand of the volume of equation \eqref{eq:rdotIntegral}  in Figure \ref{fig:integrand}. Notice that as $A \rightarrow A_c$ the major contribution of the volume increasingly comes from a small region around $r_V=\frac32 m$. 

We can visualise the surfaces we have found in an Eddington-Finkelstein diagram by integrating 
numerically  \eqref{eq:vdoteq} using the result obtained from \eqref{eq:rdoteq}. This is done in Figure \ref{fig:maxGeod}. Depicted, in these coordinates, are the maximum volume surfaces for different values of $A$, starting from the same sphere on the horizon. As $A \rightarrow A_c$ the surfaces reach $r=0$ at earlier times and the volume increases. In other words, the hypersurfaces $\Sigma_v$ elongate along $r_V$ and build up volume while in that region, before ending in lower values of $v$.  

Notice that in the region $0<r<<3m/2$, $\Sigma_v$ approach the incoming null direction. This explains why the part of each $\Sigma_v$ that is close to the singularity gives no contribution to the volume. Notice that there is no direct relation between the existence of the singularity and the volume becoming large. Locally, the behaviour of the volume in a black hole spacetime is best captured by the auxilliary metric \eqref{eq:Mauxlinel}, in which no infinities are present because of the $r^4$ factor. In a sense, the volume ``does not see'' the singularity.  

\begin{figure}[t] 
 \centering
 \includegraphics[scale=0.5]{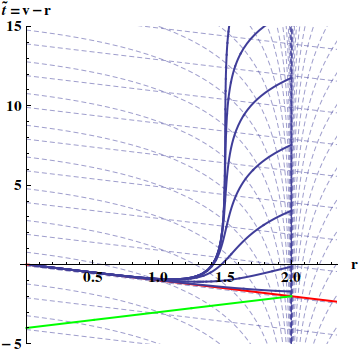}
 \caption{ \label{fig:maxGeodNullShell} The maximal volume surfaces inside a black hole formed by a collapsing object.  In red is an incoming spherical null shell that collapses and forms a singularity at $v=0$. (The region below the null shell is flat.) As (asymptotic) time passes, the interior grows. In green is the horizon.}  
\end{figure}

Finally, we return to the problem we started from: the black hole generated by a collapsing object. Since the black hole originates from a collapsed object, $v$ is determined by the collapse time. The situation is illustrated in Figure  \ref{fig:maxGeodNullShell}, where, instead of fixing $S$, we have depicted the surface of maximal volume for different $S_v$'s. As we move into the future of the black hole the volume becomes arbitrarily large. 

From the perspective of the maximal-volume spherical surfaces, the interior of a black hole is close to a cylinder of approximate radius $r_V$ which grows longer with time. 

\section{Discussion: on the validity of the generalised second law} \label{sec:Discussion}

We have observed that in a spherically symmetric context what we usually mean by ``volume inside a sphere" is the maximal proper volume of a spacelike spherically symmetric 3d hyper-surface bounded by the sphere. We have computed this volume for a spherically symmetric black hole formed by a collapsed object. We have found that the volume inside the hole is given by equation \eqref{risultato}.   

The interesting aspect of this result is that the interior volume of the black hole is large and increases with time. The interior of a black hole ``does not last long" in the sense that all timelike geodesics hit the singularity in a proper time of order $m$, but ``is very big" in the sense that a spacelike region of very large volume fits in it. This large volume increases linearly with time since the collapse.  

The interior region of a black hole keeps increasing fast, because  the horizon is an outgoing null surface. The fact that the area of this surface remains constant, which is of course due to the curvature, does not contradict the fact that an outgoing null surface encloses an increasingly large volume as time passes. From this perspective, the conformal diagram in Figure \ref{fig:Figura1} gives a pretty accurate picture of what happens.

This result might cast some doubts on a common intuition about the amount of information that a horizon may contain. An \emph{event} horizon obeys, most likely, the generalised second law \cite{Bousso1999} and therefore it makes sense to assign it its Bekenstein entropy 
\be
S=\frac{A}4
\label{entropy}
\ee
and assume that the sum of this quantity and the external entropy never decreases.  But quantum effects both inside and at the horizon are likely to make event horizons unphysical (see for instance the recent analysis of the Planck star bounce \cite{Rovelli2014,Barrau2014,Haggard2014,Barrau2014b} and references therein. A different scenario where the considerations of this paper are also pertinent has been recently put forward in \citep{Perez2014}). Because of these, the horizon of a gravitationally collapsed object might be an {\em apparent} horizon.  In these conditions, the validity of the second law of thermodynamics is obviously out of question, but the validity of the \emph{generalised} second law is far from certain. 

The information inside the hole could be recovered: after crossing the quantum region that replaces the classical singularity, it may be free to exit.  Therefore the information inside the horizon is not degraded and should not be counted as entropy. We can still associate an entropy \eqref{entropy} to an apparent horizon, because this same quantity measures the quantum field theoretical entanglement across the horizon (see \cite{Bianchi2012} and references therein). This entropy behaves precisely as a thermodynamical entropy, and is indistinguishable from the Bekenstein-Bousso entropy as long as the horizon is present; but it is a quantum von Newman entropy and, as such, nothing prevents it from decreasing when we have access to the black hole interior, which is possible if the horizon is apparent. Von Neumann entropy, of course {\em does} decrease with time, if we access more observables at later time. 

If the horizon can disappear in time, the information contained inside an horizon can exit. The second law of thermodynamics remains valid, but not its Bekenstein generalisation.  As we have shown, the interior of the black hole has plenty of room to store information \footnote{After the appearance of this paper on the arXiv, the result we point out has been developed for the spinning case by Bengtsson and Jacobson in \cite{Bengtsson2015} and for other cases by Yen Chin Ong in  \cite{YenChinOng2015}.}. \\

\section*{ACKNOWLEDGMENTS}
CR thanks Don Marolf for useful exchanges. MC thanks Alejandro Perez for extensive discussions during the past year and Goffredo Chirco, Aldo Riello and Ernesto Frodden for early discussions on this subject. MC also acknowledges support from the Educational Grants Scheme of the A.G.Leventis Foundation for the academic years 2013-2014 and 2014-2015.

\appendix

\section{Volume in Reissner-Nordstr\"{o}m} \label{app:RN}
In this appendix, we check the reasoning presented in this paper in a less trivial example. This spacetime, describing a non-rotating spherically symmetric charged black hole, can be described by \eqref{eq:iEFlinel} with
\be
f(r)=1-2M/r+Q^2/r^2. 
\ee
There are two horizons on this spacetime located at the zeros of 
$f$, given by 
\be
r_{\pm}=M\pm \sqrt{M^2-Q^2}.
\ee
The outer horizon, at $r_+$, is an event horizon and the inner horizon, at $r_-$, is an apparent horizon. It is easily calculated that between $r_-$ and $r_+$ the polynomial $r^4f(r)$ has a minimum value at 
\be
r_V=\frac14\left(3M+\sqrt{9M^2-8Q^2}\right)
\ee
Then,
\be
A_c^2>-r_V^4 f(r_V)
\ee
It is then evident that the analysis presented here is essentially the same in this spacetime. The difference is that instead of looking at the region between the singularity and the horizon, we look at the region between the two horizons. If one wishes to include the volume in the region $r<r_-$ then a different analysis must be carried out for that part since $r$ re-acquires its interpretation as a spacelike coordinate. The situation for that region would be similar to that of Minkowski space (see Appendix \ref{app:Minkowski}), i.e.\ it would yield a constant contribution that becomes negligible for large times.
\begin{figure}[t] 
 \centering
 \vspace{.5cm}
 \includegraphics[scale=0.35]{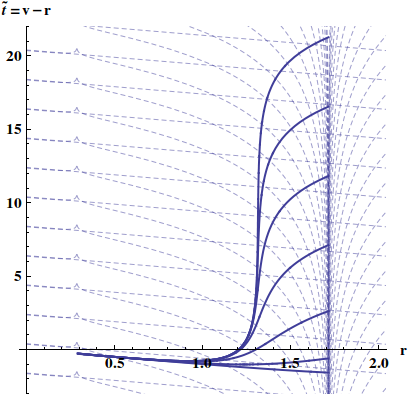}
 \caption{ \label{fig:maxGeodRN} The family $\Sigma_0$ in a Reissner-Nordstr\"{o}m black hole for $M=1$ and $Q=0.7$. The inner (apparent) horizon is at $r_-\approx 0.29$ and the outer (event) horizon at $r_+\approx 1.71$. We have taken $v_f=0$}.   
\end{figure}

\section{Volume in Kruskal} \label{app:KR}
A straightforward application of the definition of volume that we have given in the case of a Kruskal spacetime gives infinite volume for any sphere $S_v$ on the horizon.   We may obtain a finite volume by requiring $\Sigma$ to be bounded by {\em two} spheres, $S_v$ and $S_u$, one on each of the two outgoing horizons. Fixing $S_u$, the volume $V(u,v)$ of this surface clearly satisfies 
\be
        V(u,v_2)-V(u,v_1) \sim 3 \sqrt{3}\pi m^2 (v_2-v_1)
\ee
in the asymptotic region. 

\section{Volume in Minkowski space}  \label{app:Minkowski}

It is instructing to see how the definition we gave in the introduction and the results in the paper work in the trivial case of flat space. We can again define a null coordinate $v=t+r$. This is the same as setting $f(r)=1$ in the line element \eqref{eq:iEFlinel} or most of the formulas given in the paper. It is direct from the integral \eqref{eq:rdotIntegral} that the maximum volume is given by $A=0$ and that the result is the usual expression for the volume of a sphere with radius $2M$. The equations \eqref{eq:rdoteq} and \eqref{eq:vdoteq} become 
\begin{eqnarray}
\dot{r}&=& \frac{1}{r^2}\\
\dot{v}&=&\frac{1}{r^2}
\end{eqnarray}
from which we deduce that $\dd r= \dd v$. Thus, the maximal volume hypersurfaces are given by $t=v-r=constant$ as expected. Notice also that the Eddington-Finkelstein diagram of flat space is the standard depiction of Minkowski space in polar coordinates with the angular directions suppressed. That is, the vertical axis is now $\bar{t}=v-r=t$ and the EF-diagram is simply a $t$ vs $r$ plot with the usual 45 degrees causal structure.      
	
	The situation is of course very different. In this case, there is a finite maximum volume to be achieved, when $A = A_c=0$. By the characterization we gave in Section~\ref{sec:sphereVolume}, given a sphere $\Sigma$, we look for the spacelike surface $\Sigma_0$ that spans the interior of the sphere and has the largest volume. These will always be the $t=constant$ surfaces. \\

\end{document}